\documentclass[conference]{IEEEtran}
\IEEEoverridecommandlockouts
\usepackage{graphicx}
\usepackage{hyperref}
\usepackage[dvipsnames]{xcolor}
\usepackage{amsmath}
\usepackage{amssymb}
\usepackage{microtype}
\usepackage[utf8]{inputenc}
\usepackage[normalem]{ulem}
\usepackage{tabularx}
\usepackage{paralist}
\usepackage{booktabs}
\usepackage[ruled,linesnumbered,vlined]{algorithm2e}
\usepackage[noend]{algpseudocode}
\usepackage{subcaption}
\usepackage{tikz}
\usetikzlibrary{
	trees,
	matrix,
	fit
}

\RequirePackage{amsmath}

\newcommand{\sacofa}{SaCoFa\xspace}
\newcommand{\mypar}[1]{\smallskip\noindent\textbf{#1.}}
\newtheorem{definition}{Definition}

\hypersetup{
	pdftoolbar=true,        %
	pdfmenubar=true,        %
	pdffitwindow=false,     %
	pdfstartview={FitH},    %
	pdftitle={},    %
	pdfauthor={},     %
	pdfsubject={},   %
	pdfcreator={},   %
	pdfproducer={}, %
	pdfkeywords={}, %
	pdfnewwindow=true,      %
	colorlinks=true,       %
	linkcolor=NavyBlue,          %
	  citecolor=OliveGreen,        %
	  filecolor=magenta,      %
	  urlcolor=NavyBlue           %
}

\begin{document}
\title{SaCoFa: Semantics-aware Control-flow Anonymization for Process Mining}
\author{\IEEEauthorblockN{Stephan A. Fahrenkog-Petersen\IEEEauthorrefmark{1}}
	\and
	\IEEEauthorblockN{Martin Kabierski\IEEEauthorrefmark{1}}
	\and
	\IEEEauthorblockN{Fabian Rösel\IEEEauthorrefmark{1}}
	\and
	\IEEEauthorblockN{Han van der Aa\IEEEauthorrefmark{2}}
	\and	
	\IEEEauthorblockN{Matthias Weidlich\IEEEauthorrefmark{1}}
	\and
\IEEEauthorblockA{\IEEEauthorrefmark{1}\textit{Humboldt-Universität zu Berlin} \\
	Berlin, Germany \\
	\{fahrenks,bauermart,roeselfa,weidlima\}@hu-berlin.de}
\and
\IEEEauthorblockA{\IEEEauthorrefmark{2}\textit{University of Mannheim} \\
	Mannheim, Germany \\
	han@informatik.uni-mannheim.de}
}

\maketitle              %
\begin{abstract}
Privacy-preserving process mining enables the analysis of business 
processes using event logs, while giving guarantees on the protection of 
sensitive information on process stakeholders. To this end, existing approaches 
add noise to the results of queries that extract properties of an event 
log, such as the frequency distribution 
of trace variants, for analysis.
Noise insertion neglects the 
semantics of the process, though, and may generate traces not present in the 
original log. This is problematic. It 
lowers the utility of the published data and makes noise easily 
identifiable, as some traces will violate well-known semantic constraints.
In this paper, we therefore argue for privacy preservation that incorporates a 
process' semantics. For common trace-variant queries, we show how, based on the 
exponential mechanism, semantic constraints are incorporated 
to ensure differential privacy of the query result. Experiments 
demonstrate that our semantics-aware anonymization yields 
event logs of significantly higher utility than existing approaches. 
\end{abstract}


\section{Introduction}

Process mining analyses business processes based on event 
logs recorded during their execution~\cite{van2012process}. Event logs comprise 
sequences of events that reveal how a process is executed, by whom, and for 
whom. Since a log may include sensitive information about people 
involved in a process, its content is subject to privacy regulations, such as 
the GDPR~\cite{voigt2017eu}. 
Attempts to protect sensitive information through deletion 
or pseudonymisation of identifying information (e.g., names) are ineffective, 
given that the obscured information can often be 
re-identified by relating execution sequences to knowledge about the 
context of process execution~\cite{von2020quantifying}. 
To ensure that privacy regulations are still met, 
\emph{privacy-preserving process mining}~\cite{elkoumy2021privacy} strives to protect sensitive information in event logs and process mining results by ensuring that they provide well-known privacy guarantees, such as differential privacy~\cite{fahrenkrog2019providing}.

Many process mining techniques analyse a process from an abstract 
control-flow perspective, in terms of its \emph{trace-variant distribution}. 
They require information on the recorded sequences of activity 
executions, known as trace variants, and their occurrence frequencies. 
Recognising the importance of such distributions, trace-variant 
queries may be anonymised~\cite{mannhardt2019privacy}. By 
inserting noise into the variant distribution of a log, 
differential privacy, a guarantee that bounds the impact of the data of one 
individual on the query result, is ensured.  

The state-of-the-art approach to achieve differential privacy for trace-variant 
queries~\cite{mannhardt2019privacy} has an important drawback, though. 
Employing a Laplacian mechanism, it inserts noise randomly, which neglects the 
semantics of the underlying process. 
The returned trace variants may then represent behaviour that 
was never observed or, more importantly, which is clearly impossible for the 
process at hand. 
For a treatment process in a hospital, for instance, the 
anonymized distribution may include trace variants in which a patient is 
\emph{discharged} from the hospital before \emph{arriving} there.
Including such obviously incorrect sequences lowers the utility of the 
published data for process analysis, e.g., resulting in misleading models. At 
the same time, adversaries can easily recognize such trace variants as the 
result of the anonymization procedure, so that the assumed privacy guarantee no 
longer holds.
Against this background, we target the question of \emph{how to incorporate a 
process' semantics in control-flow anonymization}.

In this paper, we address the above research question with \sacofa, an approach 
for \underline{s}emantics-\underline{a}ware 
\underline{co}ntrol-\underline{f}low \underline{a}nonymization. Our idea is to 
achieve 
differential privacy of trace-variant queries based on \emph{exponential 
noise-insertion} techniques. Unlike noise insertion with the Laplacian 
mechanism, the exponential mechanism enables us to control the way noise is 
inserted, while providing the same degree of 
privacy~\cite{dwork2006calibrating}. 

Specifically, we present \sacofa as a general algorithm to achieve controlled 
noise insertion using the exponential mechanism. That includes the definition 
of a score function to assess the loss induced by the insertion of noise, which 
enables us to incorporate a process' semantics. Given the exponential runtime 
complexity of \sacofa, we further present semantics-aware optimizations for 
approximate privacy guarantees. 

Compared to the state of the art, trace-variant distributions obtained with 
\sacofa have a higher utility for process analysis and provide more robust 
privacy guarantees. 
We demonstrate these advantages in experiments with public datasets: Models 
discovered from the resulting distributions show higher F-scores 
and better generalization. Also, \sacofa introduces less obvious 
noise, as classified by anomaly detection techniques.

Below, \autoref{sec:motivation} first motivates the need for our 
work, before \autoref{sec:background} introduces background 
information. \autoref{sec:approach} presents \sacofa, 
which we evaluate in \autoref{sec:evaluation}. We review related work 
in \autoref{sec:related_work} and conclude in \autoref{sec:conclusion}.

\begin{table*}[t]
	\caption{Illustration of a trace-variant distribution, both original and 
	privatized.}
\label{tab:running_example}
	\begin{subtable}[t]{.5\linewidth}
		\centering
		\scriptsize
		\caption{Original trace-variant distribution.}
		\label{tab:example_raw_table}
		\begin{tabularx}{.99\textwidth}{l  X}
			\toprule
			Trace Variant & \#\\
			\midrule
			$\langle \mathit{Register},\mathit{Triage},\mathit{Surg.}, 
			\mathit{Release}\rangle$ & 20\\
			$\langle \mathit{Register},\mathit{Triage},\mathit{Surg.}, 
			\mathit{Antibio.},
			\mathit{Release}\rangle$ & 12\\
			$\langle \mathit{Register},\mathit{Triage},\mathit{Antibio.}, 
			\mathit{Antibio.}.
			\mathit{Release}\rangle$ 
			& 6 \\
			$\langle \mathit{Register},\mathit{Triage},\mathit{Antibio.}, 
			\mathit{Surg.}, \mathit{Release}\rangle$ 
			& 5 \\
			$\langle \mathit{Register},\mathit{Triage},\mathit{Consul.}, 
			\mathit{Release}\rangle$ 
			& 2 \\
			$\langle 
			\mathit{Register},\mathit{Triage},\mathit{Consul.},\mathit{Surg.}, 
			\mathit{Release}\rangle$ 
			& 4 \\
			\bottomrule
		\end{tabularx}
	\end{subtable}%
	\,
	\begin{subtable}[t]{.5\linewidth}
		\centering
		\scriptsize
		\caption{Privatized trace-variant distribution.}
		\label{tab:example_pertubated_table}
		\begin{tabularx}{.99\textwidth}{l  X}
			\toprule
			Trace Variant & \#\\
			\midrule
			$\langle \mathit{Register},\mathit{Triage},\mathit{Surg.}, 
			\mathit{Release}\rangle$ & 18 \\
			$\langle \mathit{Register},\mathit{Triage},\mathit{Antibio.}, 
			\mathit{Antibio.},
			\mathit{Release}\rangle$ 
			& 7 \\
			$\langle 
			\mathit{Release},\mathit{Triage},\mathit{Triage},\mathit{Surg.},\mathit{Register}\rangle$
			& 4\\
			\bottomrule
		\end{tabularx}
	\end{subtable}%
	\vspace{-1em}
\end{table*}

\section{Motivation}
\label{sec:motivation}

The state-of-the-art technique for the privatization of trace-variant 
distributions~\cite{mannhardt2019privacy} constructs a prefix tree. It 
considers prefixes of trace variants of increasing lengths and obfuscates their 
occurrence counts using the Laplacian mechanism~\cite{dwork2006calibrating}. 
Due to the exponential growth of the set of possible prefixes for a set 
of activities, infrequent prefixes are pruned to achieve an acceptable 
runtime of the algorithm. While the resulting trace-variant 
distribution is differentially private, new behaviour may have been 
\emph{introduced} to the distribution and behaviour from the original log may 
have been \emph{removed}. 
We illustrate the resulting issues using the trace-variant 
distributions in \autoref{tab:running_example}.

\mypar{Behaviour insertion problems}
The Laplacian mechanism introduces noise into a trace-variant distribution in a 
fully random manner. Any new trace variant is considered to be 
equally suitable or problematic, respectively. Depending on the underlying 
process, however, some trace variants may easily be identified as 
manipulated ones. 
For instance, the third trace variant in 
\autoref{tab:example_pertubated_table} contains a repetition of the 
$\mathit{Triage}$ activity.
Similarly, although all traces in the original log start with the prefix
$\langle \mathit{Register}, \mathit{Triage} \rangle$ and end with a 
$\mathit{Release}$ activity,
the aforementioned variant in \autoref{tab:example_pertubated_table} violates 
these patterns. 
Even without detailed knowledge about the process, an adversary immediately 
identifies this variant as artificial behaviour and omits it during an attack, 
which effectively reduces the privacy guarantee associated with the published 
query result.

\mypar{Behaviour removal problems}
The pruning strategies employed when anonymizing a trace-variant 
distribution also lead to the removal of behaviour.
In our example, the third, fourth, and fifth variants of 
\autoref{tab:example_raw_table} do not appear in 
\autoref{tab:example_pertubated_table}, i.e., they are assigned a count 
of zero. Since pruning is applied in the construction of the prefix tree, it 
may have far reaching consequences: Assigning the 
prefix $\langle \mathit{Register}, \mathit{Triage},\mathit{Consul.} \rangle$ an 
occurrence frequency below the pruning threshold implies that \emph{none} of 
the variants with this prefix will appear in the resulting distribution. In the 
worst case, this effect may materialize for the prefix $\langle 
\mathit{Register}, \mathit{Triage} \rangle$ in our example, which, arguably, 
would render the result useless for most process analyses.

\mypar{Proposed approach}
To alleviate the above issues, we argue that noise insertion shall be based on 
the exponential mechanism~\cite{mcsherry2007mechanism}. It enables us to assign 
scores to potential outputs, i.e., specific trace variants in our setting. This 
way, a prioritization of the trace variants that shall appear in the resulting 
distribution is achieved. In the remainder, we show how this idea enables us to 
incorporate the semantics of a process in the anonymization.

\section{Background}
\label{sec:background}

\mypar{Event model}
Our work focuses on the control-flow perspective of business processes. 
Therefore, we use an event model that builds upon 
a set of activities~$\mathcal{A}$. Each event in a log is assumed to correspond 
to one of these activities. 
Using~$\mathcal{E}$ to denote the universe of all events, a single execution of 
a process, i.e., a \emph{trace}, is modelled as a 
sequence of events $\xi \in \mathcal{E}^*$, such that no event can occur 
in more than one trace. An event log is a set of traces, $L \subseteq 
2^{\mathcal{E}^*}$, with $\mathcal{L}$ as the universe of event logs.
Distinct traces that indicate the same sequence of activity executions are said 
to be of  the same \emph{trace variant}, i.e., $\mathcal{A}^*$ is the universe 
of trace variants. The set of activities referenced by events in an 
event log $L$ is denoted by $\mathcal{A}(L)$.

\mypar{Trace-variant queries}
A \emph{trace-variant query} is a function $\tau(L) : \mathcal{L}  
\rightarrow \mathcal{A}^* \times \mathbb{N}$ that returns the trace-variant 
distribution of an event log $L$, i.e., it captures how often certain trace 
variants occur in $L$. 
Aside from $\tau(L)$ as the query over all trace variants, we define 
$\tau(L,v)$ as a query that returns the number of traces in $L$, for which the 
events correspond to the sequence of activities of trace variant~$v$.

\mypar{Differential privacy}
The privacy of a query can be guaranteed by fulfilling a privacy 
guarantee~\cite{wagner2018technical}. A common guarantee is \emph{differential 
privacy}~\cite{dwork2008differential}, which has been adopted by companies such 
as Apple, SAP, and Google.
The general idea behind differential privacy
is to ensure that the inclusion of the data of one individual in a certain 
dataset will not significantly change the result returned by a query over this 
data.  
In the context of our work, this means that a trace-variant query $\tau$ is 
said to preserve differential privacy, if the trace-variant distribution 
returned by query $\tau(L)$ does not significantly differ from the distribution 
returned by a query over a \emph{neighbouring} log, i.e., a log that contains 
one additional trace, $\tau(L \cup 
\{t\})$, for any trace $t \in \mathcal{E}^{*}$.

A trace-variant query ${\tau}$ that returns the actual frequency distribution, 
in general, cannot be expected to satisfy differential privacy. Hence, one 
relies on probabilistic queries $\hat{\tau}$ that approximate the true 
distribution, while satisfying the privacy guarantee. This leads to the 
following definition:

\begin{definition}[Differential Privacy]
	\label{def:ldp}
		Given a probabilistic trace-variant query~$\hat{\tau}$ and privacy 
		parameter $\epsilon \in \mathbb{R}$, query $\hat{\tau}$ provides 
		\emph{$\epsilon$-differential privacy}, if for all neighbouring pairs 
		of event logs 
		$L_1, L_2 \in \mathcal{L}$ and for all sets of possible trace-variant 
		distributions, $D \subseteq \mathcal{A}^* \times \mathbb{N}$, it holds 
		that:
		\vspace{-.5em}
		\begin{equation*}
			Pr[\hat{\tau}(L_1)\in D] \leq e^\epsilon \times 
			Pr[\hat{\tau}(L_2)\in D] 
		\end{equation*}
		
\vspace{-.5em}
\noindent
where the probability is taken over the randomness introduced by the 
		query $\hat{\tau}$.
\end{definition}

\noindent The lower the value of $\epsilon$, the stronger the privacy guarantee that is provided.
In scenarios where an individual can be part of multiple traces, the privacy 
parameter~$\epsilon$ shall be divided by the maximal number of traces related to an individual, in order to achieve the same degree of privacy.

To ensure differential privacy for a query, it is common to define a 
probabilistic query that inserts noise into the result of the original one.
This noise-insertion mechanism is generally guided by a probability 
distribution.

\mypar{Laplacian mechanism}
The Laplacian mechanism inserts noise based on a Laplacian distribution. This 
mechanism was used in~\cite{mannhardt2019privacy} to anonymize a trace-variant 
distribution. 
The impact of this mechanism generally depends on the strength of the privacy 
guarantee $\epsilon$ and the \emph{sensitivity} $\Delta f$ of some query $q$. A 
query $\hat{q}$ protected by the Laplace mechanism can formally be described as:
\begin{equation*}
	\hat{q}\gets q + Lap(\frac{\Delta f}{\epsilon})
\end{equation*}
The sensitivity $\Delta f$ depends on the maximum impact one individual can 
have on the result of query $q$. So, if $q$ is a trace-variant query ($\tau$, 
as introduced above) 
and one individual participates in at most one trace, 
the sensitivity is $\Delta f = 1$. If an individual can appear in multiple 
traces, the sensitivity is higher and more noise needs to be 
introduced to achieve $\epsilon$-differential privacy. However, in such 
scenarios, the guarantee of $\epsilon$-differential privacy may also be 
relaxed, which lowers the increase in sensitivity and still provides a 
relatively strong protection~\cite{kartal2019differential}.

When used to insert noise into a trace-variant distribution, the Laplacian 
mechanism has considerable drawbacks, see \autoref{sec:motivation}. That is, 
the probability that a certain anonymized 
trace-variant distribution is returned, only depends on the syntactic distance 
of this distribution to the actual one. 
This ignores that certain distributions are less desirable than others, 
even when they are syntactically just as different.

\mypar{Exponential mechanism}
The exponential mechanism enables a prioritization of certain query results by 
incorporating the notion of a score function into the noise insertion 
process. The score function $s$ defines some results to be more desirable than 
others for the given dataset over which the query is evaluated, i.e., the 
higher $s(d,r)$, the more desirable is the query result $r$ for the dataset 
$d$. Moreover, $\Delta s$ is the sensitivity of 
the score function, i.e., the maximum differences between scores assigned to 
the possible results for two neighbouring datasets. Then, for some query $q$ 
over dataset $d$ and privacy parameter $\epsilon$, the query $\hat{q}$ 
protected by the 
exponential mechanism is derived by choosing the result $r$ with a probability 
proportional to $e^{(\epsilon s(d,r))/(2\Delta s)}$.

The mechanism may be lifted to our case of a trace variant query. For a given 
log $L$ and a possible trace-variant distribution $D \in \mathcal{A}^* \times 
\mathbb{N}$, the score $s(L,D)$ shall capture whether $D$ is desirable in terms 
of a process' semantics, as captured by $L$. Then, 
the mechanism returns a specific trace-variant 
distribution $D$ with a probability proportional to $e^{(\epsilon s(L, 
D))/(2\Delta s)}$. 
This general idea will be exploited in our \sacofa approach, as presented in 
the next section.

\section{Semantics-aware Control-flow Anonymization}
\label{sec:approach}
This section  introduces \sacofa (\underline{s}emantics-\underline{a}ware 
\underline{co}ntrol-\underline{f}low \underline{a}nonymization) as an approach 
to retrieve the anonymized behaviour of an event log. \autoref{sec:mech_design} 
presents the general algorithm based on the exponential mechanism. 
\autoref{sec:score_function} then defines the score function, needed to 
incorporate a process' semantics. Finally, \autoref{sec:prunning} discusses 
pruning strategies for \sacofa, their computational necessity, and how the 
score function helps to decrease the negative effects of pruning.  

\subsection{The SaCoFa Algorithm}
\label{sec:mech_design}

The idea of the \sacofa algorithm is to construct a prefix tree of trace 
variants through step-wise expansion, where each step adds an activity or a dedicated end 
symbol to a branch in the tree. During this construction, prefixes are evaluated based on a \emph{score function}, which reflects
their compliance with the process' semantics, as 
captured in the original event log. Specifically, prefixes are categorized 
as \emph{harmful} or \emph{harmless}, depending on whether they violate 
semantic constraints and hence, threaten the utility of a trace-variant 
distribution. 

While harmless prefixes are always added to the tree, some harmful prefixes typically also 
need to be incorporated, to achieve differential privacy. 
To this end, we leverage the exponential mechanism, which incorporates the 
score function to assign 
lower probabilities to prefixes that induce a stronger 
violation of a process' semantics.
Hence, we are able to nudge the expansion of the tree to 
prefixes that are less harmful. In any case, all prefixes added to the tree are
assigned noisy counts. 
To cope with the exponential growth of the prefix tree, 
we also prune the tree based on these noisy counts in each step of its expansion.

In \autoref{alg:overview}, we provide the pseudo-code for our algorithm. 
It takes as input an event log $L$ and several parameters: the strength of the 
desired 
privacy guarantee $\epsilon$, an upper bound on the trace-variant length $k$,
and a pruning parameter $p$ (or two pruning parameters $p_{\mathit{harmless}}$ 
and $p_{\mathit{harmful}}$, as detailed later). It returns $\tau'(L)$, i.e., an 
anonymized trace-variant distribution. %

First, the algorithm initializes the prefix tree, represented as a set of 
prefixes~$T$
(\autoref{line:init_tree}). Next, the trace-variant distribution $d$ and the 
current prefix length $n$ are initialized 
(lines~\ref{line:init_count}-\ref{line:init_prefix}). 
Then, the prefix tree is iteratively
expanded, which will terminate when $n$ reaches the maximal prefix length $k$ (\autoref{line:loop_start}). 

\mypar{Candidate generation}
For each $n \leq k$, we expand the current tree by first generating a set of candidate prefixes.
To obtain these candidates, we select each prefix $v \in T$ that is maximal, $|v| = n-1$, and has not yet been ended, $v(|v|) \neq\ \perp$ (\autoref{line:retrieve_prefixes}). Note, the first iteration takes the empty prefix. 
Then, for each $a \in \mathcal{A}(L) \cup \{\perp\}$, i.e., for any activity or 
the end symbol $\perp$, we generate a new candidate by appending $a$ to $v$  
and add it to the  candidate set $C$ 
(lines~\ref{line:loop_activities}-\ref{line:add_candidate}).

\begin{algorithm}[t!]
	\caption{The SaCoFa Algorithm}
	\label{alg:overview}
	\footnotesize
	\SetKwInOut{Input}{input}%
	\SetKwInOut{Output}{output}%
	\Input{
		$L$, an event log;
		$\epsilon$, the privacy parameter;
		$k$, the max. prefix\\ \ length;
		$p$ ($p_{\mathit{harmless}}, 
		p_{\mathit{harmful}}$), the pruning 
		parameter(s). 
	}
	\Output{
		the result of $\tau'(L)$, an anonymized trace-variant 
		distribution.
	}
	\BlankLine
	$T \gets  \{\langle \rangle\}$\tcc*{\smaller 
		Initalize the prefix tree} 
	\label{line:init_tree}
	$d\gets \emptyset$\tcc*{\smaller Initalize the trace-variant 
	distribution} 
	\label{line:init_count}
	$n \gets 1$\tcc*{\smaller Initialize the current prefix length} 
	\label{line:init_prefix}
	
	\BlankLine
	\While(\tcc*[f]{\smaller Consider prefixes up to length $k$}){$n \leq k$} 
	{
		\label{line:loop_start}
		$C  \gets \emptyset$\tcc*{\smaller Initialize candidate set} 
		\label{line:init_candidates}
	
		\tcc{\smaller Select candidate prefixes to expand} 
		\ForEach{$v 
			\in T  \land |v| = n-1\ \land v(|v|) \neq \perp$ }{
			\label{line:retrieve_prefixes}
			\tcc{\smaller For each possible activity} 
			\ForEach{$ a \in 
				\mathcal{A}(L) \cup \{\perp\}$}{
				\label{line:loop_activities}

				\tcc{\smaller Add 
					expanded prefix to candidate set}
				$C \gets  C \cup \{v.\langle a \rangle \}$\; 
				\label{line:add_candidate}
			}
		}
	
		\tcc{\smaller Determine 
			harmless prefix candidates}
		$C_{\mathit{expand}}\gets 
		\{c\in C \mid \mathbf{score}(L,c) = 1\}$
		\label{line:score_function}\;
		\tcc{\smaller Determine harmfull prefix  candidates}
		$C_{\mathit{harm}}\gets C \setminus C_{\mathit{expand}}$
		\label{line:score_function_2}\;

		\tcc{\smaller Select prefixes; harmful prefix candidates are selected 	
		using the exponential mechanism} 
		$C_{\mathit{expand}} \gets C_{\mathit{expand}} \cup \mathit{Exp}(
		C_{\mathit{harm}}, 		
		\mathbf{score}(L,.),\epsilon)$\label{line:expand_selection}\;

		\BlankLine
		\tcc{\smaller Assign positive noisy count to prefixes} 
		\ForEach{$v \in C_{\mathit{expand}}$}{
			\label{line:expansion}
			$d(v) \gets [\tau(L,v) + 
			\mathit{Lap}(\frac{1}{\epsilon})]_{\geq 0}$ 
			\label{line:noisy_count}\;
		}
		
		\BlankLine	
		$T\gets \mathbf{prune}(T,d,p,C_{\mathit{harm}})$\tcc*{\smaller Prune 
			prefix tree} 
		\label{line:prune}
			\BlankLine	
		$n \gets n + 1$\tcc*{\smaller Increase current prefix length} 
		\label{line:increase_prefix}
	}
	\tcc{\smaller Return the distribution over all 
	prefixes that are complete or of length $k$} 
	\Return  $\{ d(v) \mid v\in T \land (v(|v|)=\ \perp \lor\ 
	|v|=k 
	)\}$\; \label{line:return}
\end{algorithm}

\mypar{Tree expansion}
The candidate prefixes in $C$ are evaluated with a \emph{score 
function} to classify them as harmless ($C_{\mathit{expand}}$) or 
harmful ($C_{\mathit{harm}}$) 
(lines~\ref{line:score_function}-\ref{line:score_function_2}). The 
definition of the score function depends on the incorporated notion of a 
process' 
semantics and will be discussed in \autoref{sec:score_function}. Here, we 
assume the score function to be applicable to 
prefixes, while the actual scoring (function $s$ in \autoref{sec:background}) 
refers to a distribution over prefixes with their frequencies all set to one. 

Employing the exponential mechanism, we determine which of the harmful 
prefixes to add to the tree by random selection 
(\autoref{line:expand_selection}). Then, the selected harmful 
prefixes, together with the 
harmless ones, expand the prefix tree (\autoref{line:expansion}). Each 
of these prefixes is assigned a noisy count, based on its number of occurrences 
in the original log and random, Laplacian noise (\autoref{line:noisy_count}). 
The latter is configured by the privacy guarantee, while the noisy count is 
enforced to be a positive value, since the decision to include the 
prefix has already been taken as part of the exponential mechanism.

\noindent
\textbf{Tree pruning.}
After expanding the prefix tree, we prune it based on the noisy counts assigned 
to trace variants (\autoref{line:prune}). A simple pruning strategy removes all 
prefixes from the tree, for which the noisy 
count is below a threshold set by parameter $p$. However, as we 
will discuss in \autoref{sec:prunning}, pruning may treat harmful and harmless 
prefixes differently (using two thresholds, $p_{\mathit{harmless}}$ 
and $p_{\mathit{harmful}}$). In general, we also favour pruning of harmful 
prefixes to avoid the removal of prefixes that conform to the 
semantics of the process at hand. 

\mypar{Result construction} Finally, the resulting trace-variant distribution is derived and returned (\autoref{line:return}). To this end, the 
counts of 
all prefixes that end with the symbol $\perp$ or that have a length of $k$ are 
considered. Intuitively, prefixes of length~$k$ may represent variants of 
traces that have not yet finished execution.

\subsection{A Semantics-aware Score Function}
\label{sec:score_function}

The \sacofa algorithm uses a score function to assess the utility loss 
associated with a prefix based on the process 
behaviour included in the original log $L$. The function is employed to 
distinguish harmless prefixes ($C_{\mathit{expand}}$) from harmful ones 
($C_{\mathit{harm}}$) 
(\autoref{line:score_function}) and in the exponential mechanism 
(\autoref{line:expand_selection}).
As such, the definition of the score function denotes a design choice that 
enables us to incorporate different notions of a process' semantics in the 
anonymization.

\begin{figure}
		\footnotesize
	\centering
\begin{tikzpicture}[level distance=1cm,
	level 1/.style={sibling distance=1.2cm}]
	\node {Register, Triage, Antibio., Surg.}
		child {node {Register}}
		child {node {Triage}}
		child {node {Surg.}}
		child {node {Antibio.}}
		child {node {Consul.}}
		child {node {Release}}
		child {node {$\perp$}};
\end{tikzpicture}
\vspace{-.6cm}
\begin{tabular}{@{\hspace{2.5em}} c @{\hspace{3.4em}} c @{\hspace{3.4em}}c 
		@{\hspace{3.4em}}c 
		@{\hspace{3.4em}}c @{\hspace{3.4em}}c @{\hspace{3.4em}} c }
	$\times$ & $\times$ & $\times$ & \checkmark & 
	$\times$ & \checkmark & $\times$\\
\end{tabular}
\vspace{.3cm}
\caption{Example for prefix expansion.}
\label{fig:prefix-extension}
\end{figure}

To exemplify this design choice, we propose a function that is based on a 
generalization of the behaviour in the original log. Specifically, we consider 
a behavioural abstraction that was proposed in the context of the 
\emph{behavioural appropriateness} measure~\cite{DBLP:journals/is/RozinatA08}. 
This behavioural abstraction defines rules between pairs of activities, 
reflecting their order and co-occurrence in a log. Specifically, given $a_1, 
a_2 \in  \mathcal{A}(L)$, 
the rules capture if $a_1$ will always, never, or 
sometimes follow (or precede) an activity $a_2$, not necessarily directly. 
As such, the set of rules encodes hidden business logic, derived from a 
log without manual intervention. 

We instantiate two score functions based on these rules, a binary and a 
continuous one. The binary instantiation classifies 
all prefixes that violate at least one rule as harmful, and all other prefixes 
as harmless. In contrast, the continuous instantiation counts the number of 
rule violations to quantify the harmfulness of a prefix. 
This degree of harmfulness is limited by 
a user-defined upper bound, since 
the sensitivity of the exponential mechanism considers the maximum impact that one trace can have on 
the score function.

As an illustration, consider the example given in 
\autoref{fig:prefix-extension}, which depicts the expansion of prefix 
$\langle \mathit{Register}, \mathit{Triage}, \mathit{Antibio.}, 
\mathit{Surg.}\rangle$, based on the example from 
\autoref{tab:running_example}. 
In the original log, 
activity $\mathit{Surg.}$ is always followed by activity $\mathit{Release}$, 
may be followed by activity $\mathit{Antibio.}$, and is never followed by the 
remaining activities. 
Respecting these behavioural rules, expansions based on the former two 
activities are considered harmless, while
those in the latter are categorized as harmful.

As mentioned above, the score function may also be defined based on other 
behavioural models. In particular, it may be grounded in other sets of 
behavioural rules, such as those presented 
in~\cite{DBLP:conf/apn/WeidlichW12,DBLP:conf/apn/PolyvyanyyWCRH14}, which are 
then instantiated for the original log to capture the semantics of the 
underlying process. Moreover, 
rules may also originate from other sources, such as textual 
documents~\cite{DBLP:journals/is/AaLR18}. 
However, deriving the rules from 
the original log ensures that trace variants in the original log are more likely to be preserved.

\subsection{Semantics-aware Pruning}
\label{sec:prunning}

To achieve differential privacy, the number of prefixes to 
be considered in the \sacofa algorithm grows exponentially in the prefix 
length. Consequently, we incorporate pruning of trace variants to achieve 
tractability, as detailed below.

\mypar{The need for generalization} 
Pruning comes with the risk of removing prefixes (and thus trace variants) that 
are common in the original log, 
which reduces  the utility of the 
anonymized trace-variant distribution. 
However, unlike log anonymization with 
the Laplacian mechanism, our approach 
supports a differentiation between prefixes that are 
harmful and harmless for an anonymized distribution. 
Therefore, we can limit pruning to harmful prefixes.
This way, the overall 
number of pruned prefixes is reduced, but harmless prefixes are always preserved, even when their noisy count is below the pruning parameter $p$.

A lower number of pruned prefixes, in general, also reduces privacy 
degradation. 
However, when pruning solely harmful prefixes, there is a risk to violate the 
required differential privacy guarantee. That is, if harmful prefixes are 
characterized based on their absence in the original log, 
the following may happen: For two neighbouring event 
logs, that differ by a trace of a variant that appears only in one of the logs, 
the anonymized variant-distributions may enable the identification of the 
respective trace. To avoid such situations, we employ a pruning strategy that 
incorporates behavioural generalization. %

\mypar{Rule-based pruning}
By employing the abstraction underlying the behavioural appropriateness measure 
to identify harmful prefixes for pruning, we avoid to reveal the difference 
between two neighbouring event logs. Due to 
the implied behavioural generalization, a 
trace representing a difference between two logs may also 
induce a change in the respective rule sets. The changed rules 
potentially allow for more behaviour, i.e., they increase the set of harmless 
prefixes. Hence, the anonymized trace-variant distributions of 
neighbouring logs may differ by \emph{multiple} trace variants, instead of just 
a single one. 

For illustration, consider a log $L_1$ containing only traces that represent 
variants from \autoref{tab:example_raw_table}. Let $L_2=L_1 \cup \{t\}$ be a 
neighbouring log, where $t$ is a trace of the variant $\langle 
\mathit{Register}, \mathit{Antibio.}, \mathit{Release}\rangle$. Comparing the 
rule sets of both logs, trace $t$ adds the rule that $\mathit{Register}$ is 
sometimes followed by $\mathit{Antibio.}$ Hence, the \sacofa algorithm would 
consider the prefix $\langle \mathit{Register}, \mathit{Antibio.}\rangle$ as 
harmless when anonymizing $L_2$, whereas it would be harmful regarding~$L_1$. 
For~$L_2$, further prefixes would then be derived and considered as harmless, 
e.g., $\langle 
\mathit{Register}, \mathit{Antibio.}, \mathit{Release}\rangle$ and
$\langle \mathit{Register}, \mathit{Antibio.}, \mathit{Surg.}, 
\mathit{Release}\rangle$. Hence, the distributions derived for 
the logs will differ by more than one trace variant.

Therefore, pruning only harmful prefixes requires that a single trace either leads to multiple trace variants to be considered as harmless, or none at all.
In practice, this may 
not be the case, which is why we relax the pruning strategy, as follows. We 
introduce $p_{\mathit{harmless}}$ and $p_{\mathit{harmful}}$ as separate 
pruning thresholds for harmless and harmful prefixes, respectively. By setting 
$1<p_{\mathit{harmless}}<p_{\mathit{harmful}}$, we favour pruning of harmful 
prefixes. Yet, by pruning also some harmless prefixes, we ensure that information 
on the existence of a single trace variant is not disclosed, even if the above 
requirement is not met. Also, the two aforementioned extreme scenarios could be 
configured accordingly, i.e., pruning only harmful traces 
($p_{\mathit{harmless}}=1$ and $p_{\mathit{harmful}}>1$) or pruning all 
prefixes that introduce new behaviour ($p_{\mathit{harmless}}=1$ and 
$p_{\mathit{harmful}}=\infty$).

\section{Evaluation}
\label{sec:evaluation}

In this section, we investigate if control-flow anonymization with \sacofa provides 
higher utility for process discovery than the state of the art. 
We first review the used datasets (\autoref{sec:datasets}) and our experimental 
setup (\autoref{sec:setup}). We then present our experimental results 
(\autoref{sec:results}), before we close with a qualitative discussion of 
the approach (\autoref{sec:discussion}).

\subsection{Dataset}
\label{sec:datasets}

We use three real-world event logs as a basis for our experiments, of which 
some characteristics are listed in \autoref{tab:datasets}.
We selected these logs since they differ in their size and complexity. The 
\emph{Traffic Fines} log contains data on a very structured process, with just 
231 variants over a total of 150,370 traces. In contrast, the 
\emph{Sepsis} log captures an unstructured hospital-treatment process, 
containing 846 variants of which the vast majority occurred just 
once. Finally, the CoSeLoG event log provides a middle ground, with a 
semi-structured process that consists of 116 variants over 1,434 cases.

\begin{table}[h!]
	\vspace{-.2em}
	\caption{Descriptive statistics for the event logs.}
	\label{tab:datasets}
	\footnotesize
	\centering
	\begin{tabular}{l @{\hspace{1.5em}} r @{\hspace{1.5em}} r @{\hspace{1.5em}} 
	r 
	@{\hspace{1.5em}} r}
		\toprule
		Event Log & \# Events & \# Activities &  \# Cases & \# Variants \\
		\midrule
			CoSeLoG~\cite{Buijs2014}& 8,577& 27 & 1,434 & 116\\
		Sepsis~\cite{mannhardt2016sepsis} &15,214 & 16& 1,050& 846\\
		Traffic Fines~\cite{trafficfines} & 561,470&11 & 150,370& 231\\	
		\bottomrule
	\end{tabular}
\end{table}

\subsection{Experimental Setup}
\label{sec:setup}

\mypar{Baseline}
We evaluate our approach against 
the state-of-the-art approach by Mannhardt et 
al.~\cite{mannhardt2019privacy}, which anonymizes the result of 
trace-variant queries based on the Laplacian mechanism.

\mypar{Parameter settings}
As specified in \autoref{sec:mech_design}, \sacofa takes four parameters:
the strength of the desired privacy guarantee $\epsilon$, an upper bound on the 
trace-variant length $k$, and the pruning parameters $p_{\mathit{harmful}}$ and 
$p_{\mathit{harmless}}$.
Per event log, we set $k$ so that roughly 80-90\% of the original trace 
variants are covered. 
For each of the employed privacy guarantees, i.e., $\epsilon = \{1.0,0.1,0.01 
\}$, we explored pruning parameters starting at 2, 20, and 200, respectively, 
until a configuration was found such that the trace-variant query could be 
executed within several seconds.
Overall, this approach resulted in the parameter settings given in 
\autoref{tab:dataset_epsilon}, which we employed for our approach and, if 
applicable, for the baseline.

\begin{table}[t]
	\caption{Employed parameter settings.}
	\label{tab:dataset_epsilon}
	\centering
	\footnotesize
	\begin{tabular}{l  @{\hspace{1.5em}} r @{\hspace{1.5em}} r 
			@{\hspace{1.5em}} r @{\hspace{1.5em}} r}
		\toprule
		\textbf{Log} & 
		$\epsilon$ & $k$ & $p_{\mathit{harmful}}$  & 
		\textbf{$p_{\mathit{harmless}}$} \\ 
		\midrule 
		&$1.0$& 10& 3&/\\
		CoSeLoG & $0.1$& 10 & 25& 22 \\
		& $0.01$ &  10&  220 &  200 \\
		
		\midrule 
		& $1.0$&23 &4 & / \\
		Sepsis & $0.1$&  23 & 20& 15  \\
		& $0.01$ & 23& 190 & 150 \\
		
		\midrule 
		& $1.0$&9 & 2& /\\ 
		Traffic Fines  & $0.1$ &  9 &20 &15  \\
		& $0.01$ & 9&  150& 120 \\
		
		\bottomrule			
	\end{tabular}
		 \vspace{-0.0cm}
\end{table}

\mypar{Evaluation measures}
To quantify the efficacy of our work, we assess the utility of process models discovered on the basis of the anonymized trace-variant distributions generated by \sacofa and the baseline.
For this discovery, 
 we employ the Inductive Miner Infrequent~\cite{LeemansFA13} 
with the default noise threshold of 20\%.
Then, 
we determine the utility of a discovered model by measuring its $F$-score in 
relation to the original event log, i.e., the harmonic mean of the 
\emph{fitness}~\cite{berti2019reviving} and  
\emph{precision}~\cite{munoz2010fresh}.
Also, we evaluate the \emph{generalization}~\cite{buijs2014quality} of 
the process models discovered from the anonymized event logs.
 
As a second evaluation dimension, we measure the fraction of easily-recognizable %
 noise 
introduced into the anonymized event logs. To this end, we apply a standard anomaly 
detection technique, which employs isolation forests~\cite{liu2008isolation}, to 
the anonymized event logs. We train the model on the original log, before using 
it to detect  anomalous traces in the anonymized logs. 
As features in the learning process, we use a binary encoding of the 
activities, 
signalling if they are present in a trace. Moreover, we also encode the 
presence of directly-follows relations in a trace, with a binary encoding.

\mypar{Implementation}
To conduct our experiments, we implemented \sacofa in Python. 
The source code is available  on 
GitHub\footnote{\url{https://github.com/samadeusfp/SaCoFa}} under the MIT 
license.  Furthermore, we used PM4Py's~\cite{berti2019process} implementation 
of the Inductive Miner and the evaluation measures. The implementation of the 
isolation forest is available in 
scikit-learn.\footnote{\url{https://scikit-learn.org/stable/}}

\mypar{Repetitions}
To account for the non-deterministic nature of the algorithms,
we perform 10 repetitions of all experiments.
In the remainder, we report on the median and the  bounds for the upper and lower quartile, using box plots.

\begin{figure}[t]
	\centering
	\includegraphics[width=\linewidth]{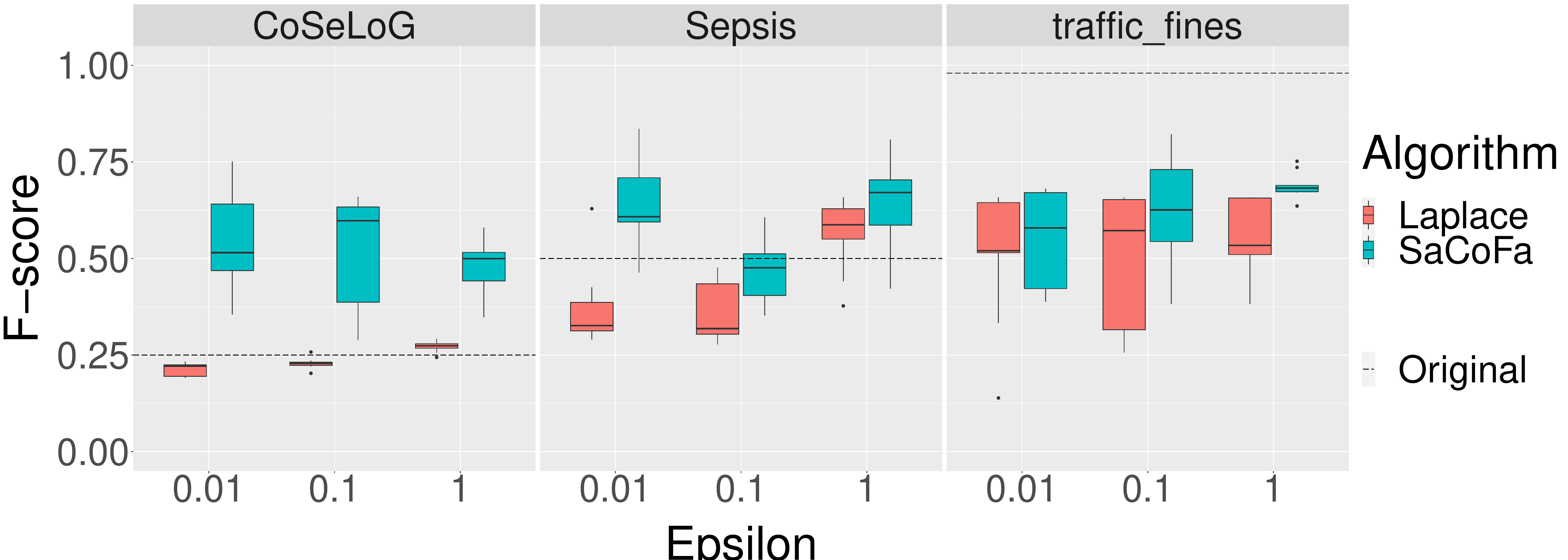}
	\caption{F-score of discovered process models.}	
	\label{fig:fscore}
	\vspace{-0.2cm}
\end{figure}

\begin{figure}[t]
	\begin{subfigure}{0.5\textwidth}
		\includegraphics[width=\linewidth]{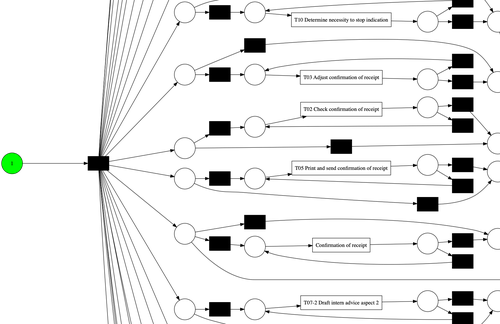}
		\caption{Laplacian baseline.}	
		\label{fig:im_laplace}
	\end{subfigure}
	\begin{subfigure}{0.5\textwidth}
		\includegraphics[clip,trim=0cm 3cm 0cm 
		0cm,width=\linewidth]{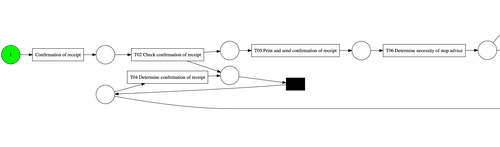}
		\caption{SaCoFa approach.}
		\label{fig:im_sacofa}
	\end{subfigure}
	\caption{Process models obtained for anonymized versions of CoSeLoG 
		($\epsilon=0.01$).}
	\label{fig:pm_models}
	\vspace{-0.2cm}
\end{figure}

\subsection{Results}
\label{sec:results}

\autoref{fig:fscore} depicts the $F$-scores
of the process models generated 
by \sacofa, as well as the Laplacian baseline.  
As shown, \sacofa 
outperforms the baseline considerably, being on par only for the setting with the strongest 
privacy guarantee ($\epsilon=0.01$) and the most structured process (Traffic 
Fines). 

In particular, we observe major improvements for the two less-structured 
processes, which have more activities and longer traces, resulting in significantly higher $F$-scores obtained using \sacofa, while providing the same privacy guarantee. 
Furthermore, our results also illustrate that, sometimes, the anonymized event 
logs lead to higher F-scores than the original log. 
The reason being that the Inductive Miner guarantees the generation of a 
fitting model, which may result in very low precision values. 
If an anonymized log contains less behaviour, above the threshold adopted by 
the discovery algorithm to filter noise, the model becomes more compact. It 
then shows higher precision and, therefore, also a higher F-score.

To further illustrate the above results, 
\autoref{fig:pm_models} shows excerpts of the process models 
obtained for the CoSeLoG process under the strongest privacy guarantee 
($\epsilon=0.01$).
Here, \autoref{fig:im_laplace} shows part of the model discovered from the log anonymized 
with the Laplacian 
baseline, while \autoref{fig:im_sacofa} is based on \sacofa.
As seen, the process model generated with \sacofa is much more structured. It starts 
with a sequence of activities that, notably, is also the same in the process 
model generated from the original event log. 
In contrast, the model in \autoref{fig:im_laplace} is very unstructured and strays far from the original process: nearly all activities can start a trace, be skipped, 
or executed multiple times. 

\begin{figure}[t]	
	\centering
	\includegraphics[width=\linewidth]{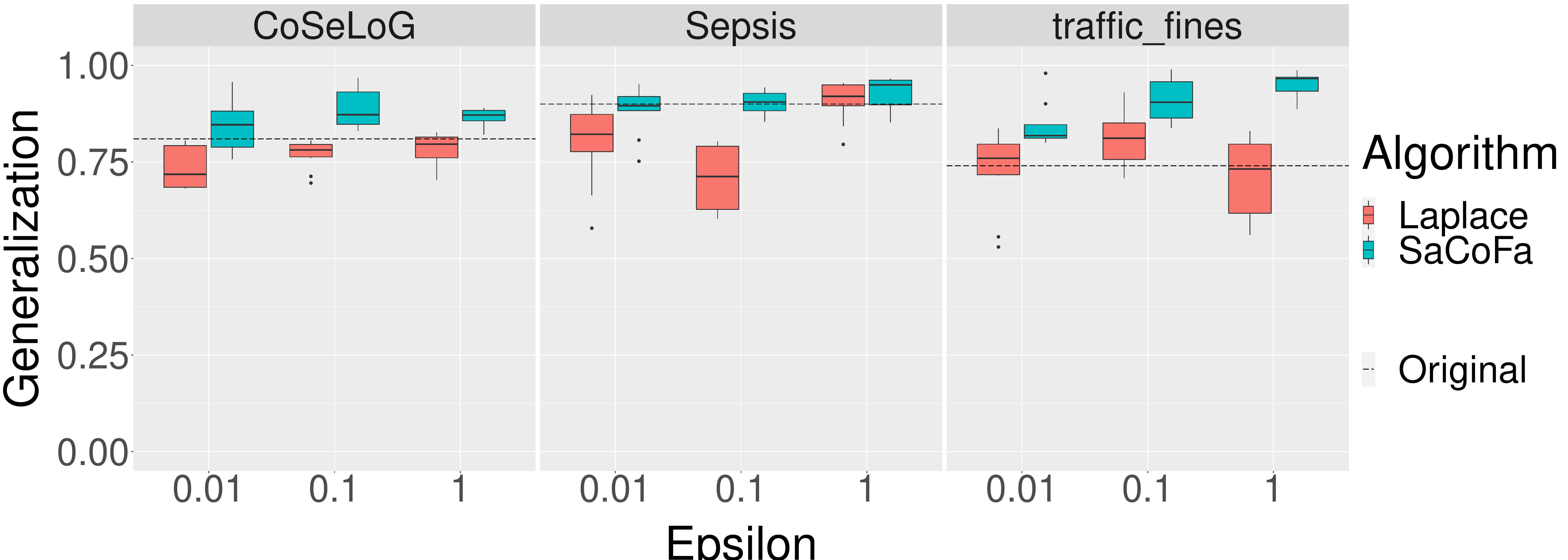}
	\caption{Generalization of discovered process models.}
	\label{fig:generalization}
	    \vspace{-0.0cm}
\end{figure}

Next, we turn to an assessment of the generalization of the obtained models. 
As illustrated in \autoref{fig:generalization}, the models 
generated based on the logs derived with \sacofa are more general, i.e., they 
abstract more from the behaviour represented in the event log. Combined with 
the results for the $F$-score, we conclude that the logs anonymized with 
\sacofa indeed have a higher utility for process discovery.

\begin{figure}[t]
	\centering
	\includegraphics[width=\linewidth]{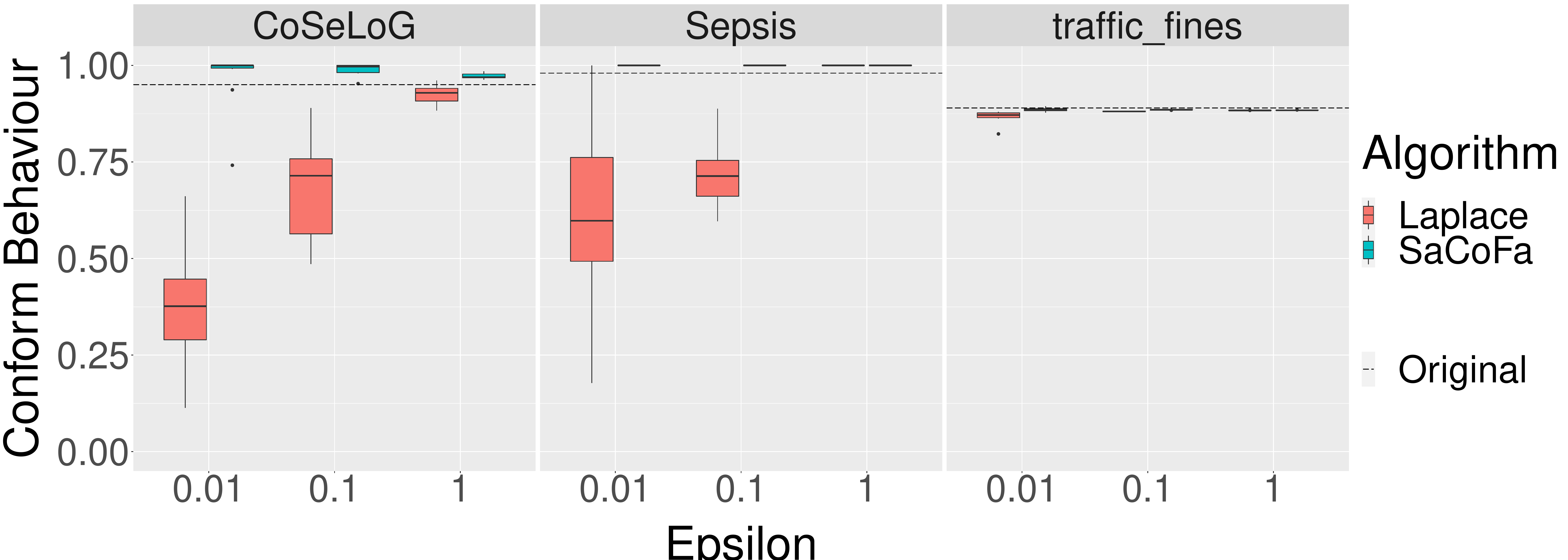}
	\caption{Relative frequency of normal behaviour in  event logs.}
	\label{fig:noiseintrusion}
	    \vspace{-0.2cm}
\end{figure}

After investigating the utility for process discovery, we turn to our second 
evaluation perspective, the presence of easily-recognizable %
 noise.
In \autoref{fig:noiseintrusion}, we show the percentage of behaviour that is 
classified as normal behaviour by the aforementioned anomaly detection 
technique.
While \sacofa and the baseline both achieve good results for the Traffic Fines 
dataset, there is a clear trend for the other two logs: 
the baseline produces much more noise, directly recognizable as anomalous. Therefore, the traces introduced by \sacofa are more in line with the original process' behaviour.
 
\begin{figure}[t]
 	\centering
	\includegraphics[width=\linewidth]{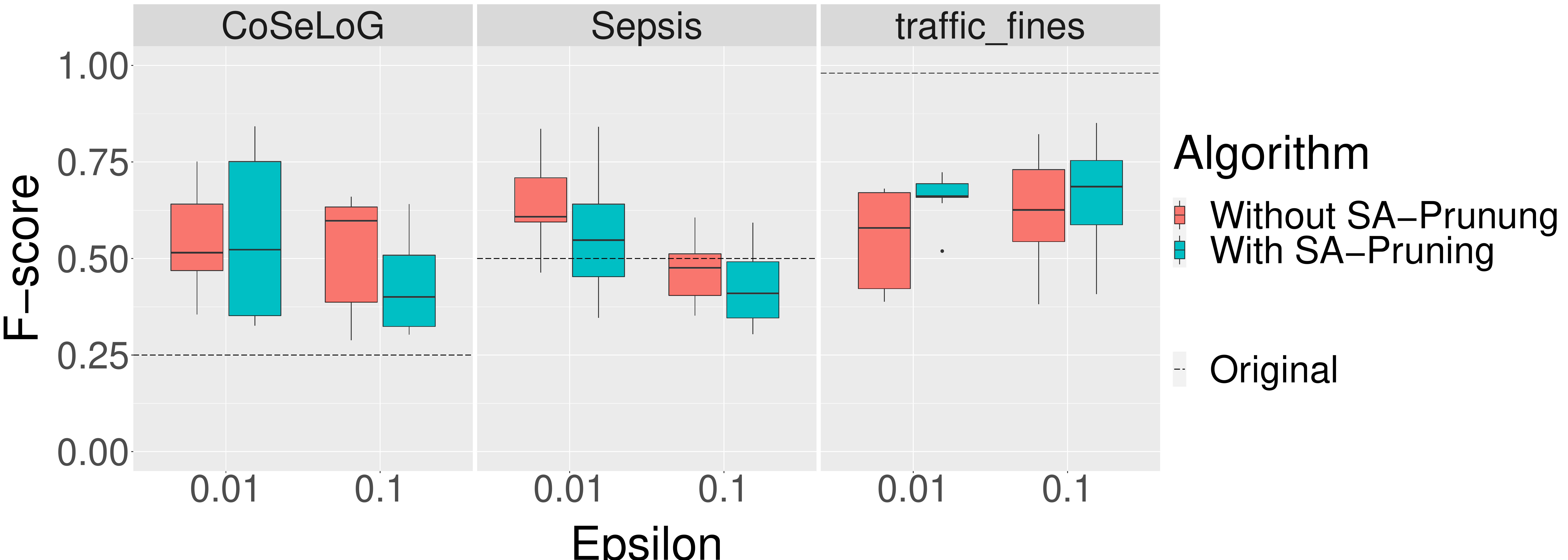}
	\caption{F-score for configurations with and without pruning.}	
	\label{fig:fscore_pruning}
		 \vspace{-0.2cm}
\end{figure}

Finally, we investigate the effects of semantics-aware pruning, as all 
previously shown result have been obtained with regular pruning. 
\autoref{fig:fscore_pruning} shows the $F$-score of models discovered from logs 
anonymized 
with and without pruning. 
Overall, 
 semantics-aware pruning turns out to only be 
beneficial for the Traffic Fines log, which is the most structured one. 
For the less-structured logs, the $F$-score actually decreases in 
comparison to the approach without pruning.
We attribute this 
observation to the significance of the rules used to separate 
harmful and harmless prefixes. Apparently, they are not always sophisticated 
enough to compensate for the additional variance introduced to the 
trace-variant distribution.

\subsection{Discussion}
\label{sec:discussion}

\mypar{Runtime aspects} 
As mentioned in \autoref{sec:setup}, parameter settings must be carefully selected in order for trace-variant queries to complete in a reasonable time.
Specifically, 
we observed that the anonymization procedure either terminated within seconds, or not all, for both \sacofa and the baseline.
This reveals that there is a clear point 
when the prefix growth makes the trace-variant
query intractable. 
So far this point is determined by step-wise altering the maximal variant 
length ($k$) and pruning parameters for a given $\epsilon$.
Nevertheless,
we observe that  \sacofa can compute results for lower pruning thresholds than the Laplacian baseline.
However, to ensure a fair comparison, we used the same pruning parameters for all 
mechanisms in our experiments.

\mypar{Non-binary score functions} 
Beyond determining if a prefix is harmful or not, the behavioural appropriateness-based score can be used to quantify its degree of harmfulness.
However, the sensitivity of the exponential mechanism depends on the maximal impact that a single case can have on the score function, i.e., on the function's maximal value  (see \autoref{sec:background}).
Therefore, if we define a score function that quantifies harmfulness in, e.g., the range $[0,3]$, 
the query's sensitivity would be $\Delta f = 3$, instead of $\Delta f = 1$ for a binary assessment.
Since the exponential mechanism needs to insert more noise for a higher sensitivity, 
the benefit obtained from quantifying harmfulness in a non-binary manner, must outweigh the increased sensitivity that comes with it.
While this did not appear to be the case in our current experiments, 
we believe that this approach could still work for more sophisticated score functions, tailored to the specifics of the process at hand. 
In any case, it is important to consider this trade-off and take it into account when choosing the right pruning parameter values.

\section{Related Work}
\label{sec:related_work}

We introduced \sacofa as an approach for control-flow anonymization that 
answers trace-variant queries, while guaranteeing differential privacy.  
The state of the art to derive a trace-variant distribution, and the related 
directly-follows graph, under differential privacy, uses the 
Laplacian mechanism~\cite{mannhardt2019privacy}. As discussed, this neglects 
a process' semantics, leading to potentially low data utility and noise that can 
be easily recognized. For trace-variant queries, Elkoumy et 
al.~\cite{elkoumy2020privacy} further studied the relation between utility and 
risk, while the PRIPEL framework~\cite{fahrenkrog2020pripel} uses 
trace-variant queries as a basis for privacy-preserving event log publishing.

Beyond differential privacy, the anonymization of event logs based on other 
privacy guarantees was studied. PRETSA~\cite{fahrenkrog2019pretsa} sanitizes 
event logs to ensure $k$-anonymity 
and $t$-closeness,
which are guarantees based on the idea of 
grouping similar cases together. Furthermore, a process mining-specific 
extension of $k$-anonymity, called TLKC, was introduced 
in~\cite{rafiei2021tlkc}. Previous work focused on improving the
utility of these techniques through feature learning-based distance metrics~\cite{rosel2021distance}.
Another group-based approach was introduced by Batista et 
al.~\cite{batista2021uniformization}, based on the uniformization of events 
within a group of individuals. The issue of continuously publishing anonymized event logs
was studied in~\cite{rafiei2021privacy}.

Approaching privacy preservation from the viewpoint of the analysis techniques 
used in process mining, it was shown how 
multi-party computation enables the construction of 
process models based on inter-organizational processes, without sharing the 
data between parties~\cite{elkoumy2020secure}. 
Other approaches target privacy-aware role mining~\cite{rafiei2019mining} and the establishment of privacy-aware process performance indicators through the enforcement of differential privacy~\cite{kabierski2021privacy}.

\section{Conclusion}
\label{sec:conclusion}

Targeting control-flow anonymization for event logs, we introduced the \sacofa 
approach to answer trace-variant queries. 
It is based on a prefix tree construction and the exponential mechanism. Unlike 
state-of-the-art 
techniques that leverage the Laplacian mechanism and, hence, introduce noise 
randomly to achieve 
differential privacy, \sacofa incorporates the 
semantics of the underlying process when inserting noise, achieving the same 
privacy guarantee. Specifically, we introduced a score function that 
differentiates prefixes as being harmful or harmless, thereby incorporating a 
process' semantics in the anonymization. We further showed how an assessment of 
the harmfulness of prefixes may also guide pruning decisions in the prefix tree 
construction in order to achieve tractability. 

Our evaluation experiments highlight that 
process models generated based on control-flow behaviour anonymized with \sacofa
have higher utility than those obtained with the state of the art. At the same 
time, the models are more general and, hence, abstract better from the 
behaviour represented in the event log. Moreover, we also showed that 
\sacofa introduces less noise that is directly labelled as anomalous 
compared to the state of the art.

\bibliography{references}
\bibliographystyle{IEEEtran}

\end{document}